\documentclass[letterpaper,12pt]{article}
\usepackage{lscape}
\usepackage{amsmath}
\usepackage{amssymb}
\usepackage[dvips]{graphicx,epsfig}
\usepackage{psfrag}
\title{Dirac quasinormal modes of $D$-dimensional de Sitter spacetime}

\author{A.\ L\'opez-Ortega\thanks{Electronic address: alfredo@fis.cinvestav.mx} \\ Facultad de Ciencias, Universidad de Colima \\ Bernal Diaz del Castillo 340\\ Colima, Colima, M\'exico.\\ and \\ Departamento de F\'{\i}sica, CINVESTAV IPN\\ Apdo. Postal 14-740, 07000 M\'exico D. F., M\'exico.  }

\begin{document}

\maketitle

\begin{abstract}

We find exact solutions to the Dirac equation in $D$-dimensional de Sitter spacetime. Using these solutions we analytically calculate the de Sitter quasinormal (QN) frequencies of the Dirac field. For the massive Dirac field this computation is similar to that previously published for massive fields of half-integer spin moving in four dimensions. However to calculate the QN frequencies of the massless Dirac field we must use distinct methods in odd and even dimensions, therefore the computation is different from that already known for other massless fields of integer spin.

\end{abstract}



\section{Introduction}
\label{Section 1}

The recent studies on the propagation of classical fields in $D$-dimensional ($D \geq 4$) spacetimes are mainly motivated by the Brane World scenario in String Theory and the scrutiny of the higher dimensional features of General Relativity. Also, the dS-CFT and AdS-CFT correspondences have motivated the study of classical fields moving in spacetimes with dimension different from four, generally in asymptotically de Sitter or anti-de Sitter backgrounds (see Ref.\ \cite{Kanti:2004nr} for a review and for some references on this topic see \cite{Frolov:2002xf}--\cite{Hod:1998vk}). 

As is well known, using classical fields as a probe we can explore the properties of a spacetime. An essential part of this analysis is the computation of the quasinormal (QN) frequencies, because they depend on the parameters that characterize the spacetime, for example, the mass, charge, and angular momentum \cite{Kokkotas:1999bd}. The quasinormal modes (QNMs) were first computed for classical fields propagating in four dimensional black holes, because it is expected that the QN frequencies of the gravitational perturbations can be detected by the gravitational wave observatories and these are also useful to analyze the linear stability of the black holes \cite{Kokkotas:1999bd}.

Recently these modes have been helpful in other research lines as the AdS-CFT and dS-CFT correspondences of String theory and the thermodynamics of black holes \cite{Abdalla:2002hg}, \cite{Horowitz:1999jd}, \cite{Hod:1998vk}. Also, as in four dimensions, the QNMs are useful tools to study the physical features of the $D$-dimensional backgrounds. These facts have motivated the computation of the QN frequencies for different fields moving in $D$-dimensional backgrounds \cite{Motl:2003cd}--\cite{Horowitz:1999jd}.

A simple solution to the field equations of General Relativity is the de Sitter background \cite{Spradlin:2001pw}. Owing to its simplicity the propagation of classical fields in this spacetime has been studied, see for example \cite{Natario:2004jd}--\cite{Abdalla:2002hg}, \cite{Spradlin:2001pw}--\cite{Cotaescu:1998ay} and references therein. Its QNMs are defined as the solutions to the equations of motion for the classical fields that satisfy the boundary conditions \cite{Lopez-Ortega:2006my}, \cite{Choudhury:2003wd}, \cite{Lopez-Ortega:2006ig}: (i) the field is regular at $r=0$; (ii) the field is purely outgoing near the cosmological horizon. 

The exactly solvable systems are usually limits of more realistic systems and allow us to study in detail some properties of a physical process and test some methods which can be used to analyze more complicated systems. Thus they are powerful tools in many research lines. Therefore we expect that the examples of exactly computed QN frequencies for $D$-dimensional spacetimes may play an important role in future research. For some examples of $D$-dimensional backgrounds whose QN frequencies have been exactly computed see Refs.\ \cite{Aros:2002te}--\cite{Abdalla:2002hg}.

The exact computation of the QN frequencies for the de Sitter spacetime has attracted much attention and in Refs.\ \cite{Natario:2004jd}--\cite{Abdalla:2002hg}, \cite{Choudhury:2003wd}--\cite{Myung:2003ki} have been exactly calculated for several fields and different spacetime dimensions. We expect that these values of the QN frequencies can be useful to explore the physical properties of the de Sitter spacetime. Our main objective here is to provide an additional field for which its de Sitter QN frequencies can be exactly calculated.

Nevertheless, notice that in Refs.\ \cite{Natario:2004jd}--\cite{Abdalla:2002hg}, \cite{Choudhury:2003wd}--\cite{Myung:2003ki} are reported different conclusions on the existence of these QN frequencies for the massless fields. For example, in Refs.\ \cite{Choudhury:2003wd}, \cite{Brady:1999wd} is asserted that de Sitter QN frequencies of the massless scalar field do not exist in four dimensions. Also, Natario and Schiappa in Ref.\ \cite{Natario:2004jd} claim that for the gravitational perturbations the QN frequencies are well defined only in odd spacetime dimensions. In Refs.\ \cite{Lopez-Ortega:2006my}, \cite{Lopez-Ortega:2006ig} we showed that the de Sitter QN frequencies are well defined for electromagnetic and gravitational perturbations in $D \geq 4$ dimensions. Furthermore, in Ref.\ \cite{Myung:2003ki} Kim and Myung affirmed that the de Sitter QN frequencies do not exist; however in the latter reference they only search real frequencies and, as is well known, the QN frequencies are complex numbers.\footnote{It is convenient to note that in Ref.\ \cite{Myung:2003ki} Kim and Myung impose slightly different boundary conditions on the fields.} 

For the massive scalar field in Ref.\ \cite{Du:2004jt} Du \textit{et al.\ }asserted that this field has well defined de Sitter QN frequencies only if the mass of the field satisfy some conditions. Using a different definition for the radial flux of the scalar field to that of \cite{Du:2004jt}, we showed in Ref.\ \cite{Lopez-Ortega:2006my} that the QN frequencies are well defined without any restriction on the mass of the Klein-Gordon field whenever the test field approximation holds. Furthermore in Ref.\ \cite{Abdalla:2002hg} were found well defined de Sitter QN frequencies for the scalar field in $D$-dimensions but Abdalla \textit{et al.\ }did not analyze them in detail (see Ref.\ \cite{Lopez-Ortega:2006my} for more details).

As is well known, sometimes the Dirac field behaves in a different way than the integer spin fields when they propagate in a curved background. Thus for the Dirac field is interesting to compute its QN frequencies in de Sitter spacetime. In three and four dimensions these frequencies were already calculated in Refs.\ \cite{Du:2004jt}, \cite{Lopez-Ortega:2006ig}. Nevertheless, in the previous references for a Dirac field moving in $D$-dimensional de Sitter background, for $D \geq 5$, the de Sitter QN frequencies were not computed. 

Taking into account the recent results on the separation of variables for the Dirac equation in spherical symmetric $D$-dimensional backgrounds \cite{Cotaescu:2003be}, \cite{Das:1996we}, \cite{Gibbons:1993hg} we find exact solutions for this equation in $D$-dimensional de Sitter spacetime. Using these solutions we calculate the QN frequencies of the massive and massless Dirac field and therefore we extend the results of the previous references, filling a gap in the literature. We compute separately the QN frequencies for the massless and massive Dirac field in order to study in detail the former case, because for the massless fields there are different results on the existence of the de Sitter QN frequencies, as we already commented. 

We organize this paper as follows. In Sect.\ \ref{Section 2} we briefly describe the main results of Ref.\ \cite{Cotaescu:2003be} on the reduction of the Dirac equation to a system of ordinary differential equations when the $D$-dimensional background spacetime is spherically symmetric. In Sect.\ \ref{Section 3} we find exact solutions to the radial equations of the previous section in $D$-dimensional de Sitter spacetime and using them we calculate the QN frequencies of the Dirac field; we first study the massless case and then the massive case. In Sect.\ \ref{Section 4} we make a brief discussion on the results obtained.

In Appendix \ref{Appendix 1} we show that if the background spacetime is $D$-di\-men\-sion\-al de Sitter spacetime, then the system of differential equations for the radial functions of the massless Dirac field given in Ref.\ \cite{Das:1996we} reduces to the system that we solve in Sect.\  \ref{Section 3}. In Appendix \ref{Appendix 2} we find the effective potentials of the Schr\"odinger type equations for the Dirac field moving in $D$-dimensional de Sitter spacetime and study some related facts. In Appendix \ref{Appendix 3}  for the de Sitter QN frequencies of integer spin fields we explain and use an alternative method of computation.

\section{Dirac equation in $D$-dimensional de Sitter spacetime}
\label{Section 2}

Using a Cartesian gauge, Cotaescu in Ref.\ \cite{Cotaescu:2003be} proved that in $D$-di\-men\-sion\-al spherically symmetric spacetime the Dirac equation reduces to a system of differential equations for the radial functions. In this section we make a brief summary of the results presented in that reference where more details can be consulted (see also \cite{Cotaescu:1998ay}).

To simplify the Dirac equation, in Ref.\ \cite{Cotaescu:2003be} Cotaescu write the metric of a $D$-di\-men\-sion\-al spherically symmetric spacetime in the form\footnote{In this manuscript we adopt a notation slightly different from that of Ref.\ \cite{Cotaescu:2003be}.} 
\begin{equation} \label{e: metric cotaescu}
{\rm d} s^2 = \mathcal{W}^2 {\rm d} t^2 - \frac{\mathcal{W}^2}{\mathcal{U}^2} {\rm d}y^2 - \frac{\mathcal{W}^2}{\mathcal{V}^2} y^2 {\rm d} \Omega^2_p,
\end{equation} 
where $\mathcal{U}$, $\mathcal{V}$, and $\mathcal{W}$ are functions of $y$, ${\rm d} \Omega^2_p$ is the metric of the $p$-dimensional unit sphere and $D=p+2$. Using the form (\ref{e: metric cotaescu}) for the metric of the background spacetime, in Ref.\ \cite{Cotaescu:2003be} is shown that the equation of motion for a Dirac field of mass $M$ simplifies to 
\begin{equation} \label{e: Dirac matrix form}
\left( \begin{array}{cc} 
M \mathcal{W} & \mathcal{U} \frac{{\rm d}}{{\rm d} y} + \kappa \frac{\mathcal{V}}{y} \\
\qquad & \qquad \\
- \mathcal{U} \frac{{\rm d}}{{\rm d} y} + \kappa \frac{\mathcal{V}}{y} & - M \mathcal{W} 
\end{array} \right)  \left( \begin{array}{c} f_{\omega, \kappa}^{(1)} (y)\\ \\ f_{\omega, \kappa}^{(2)}(y) \end{array} \right)
= \omega \left( \begin{array}{c} f_{\omega, \kappa}^{(1)}(y) \\ \\ f_{\omega, \kappa}^{(2)}(y) 
\end{array} \right),
\end{equation} 
if we use a Cartesian gauge and make the following ansatz for the Dirac field
\begin{equation} \label{e: ansatz separation 1}
\Psi_{\omega, \kappa} = y^{-p/2} \textrm{e}^{-i \omega t} \left( \begin{array}{c} 
f_{\omega, \kappa}^{(1)}(y) \phi_\kappa (\Omega_p) \\ \quad \\ i f_{\omega, \kappa}^{(2)}(y) \phi_{-\kappa} (\Omega_p)
\end{array} \right),
\end{equation} 
for even $D$, and
\begin{align}\label{e: ansatz separation 2}
\Psi_{\omega, |\kappa|} &= y^{-p/2} \textrm{e}^{-i \omega t} \left[f_{\omega, |\kappa|}^{(1)}(y) \phi_{|\kappa|} (\Omega_p) + i  f_{\omega, -|\kappa|}^{(2)}(y) \phi_{-|\kappa|} (\Omega_p) \right], \nonumber \\
\Psi_{\omega, -|\kappa|} &= y^{-p/2} \textrm{e}^{-i \omega t} \left[f_{\omega, -|\kappa|}^{(1)}(y) \phi_{-|\kappa|} (\Omega_p) + i  f_{\omega, |\kappa|}^{(2)}(y) \phi_{-|\kappa|} (\Omega_p) \right],
\end{align} 
for odd $D$. In formulas (\ref{e: ansatz separation 1}) and (\ref{e: ansatz separation 2}), $\phi_{\kappa}(\Omega_p)$ represents the angular spinors appropriate for the dimension of the spacetime under study and the quantity $\kappa$ is equal to \cite{Cotaescu:2003be}
\begin{equation}
\kappa = \pm \left(\frac{p}{2} + n \right), \qquad n=0,1,2,\dots.
\end{equation} 

In static coordinates the $D$-dimensional de Sitter metric takes the form \cite{Spradlin:2001pw}
\begin{equation} \label{e: static de Sitter}
{\rm d} s^2 = \left(1 - \frac{r^2}{L^2} \right) {\rm d} t^2 - \frac{{\rm d}r^2}{\left(1 - \frac{r^2}{L^2} \right)} - r^2 {\rm d} \Omega_p^2,
\end{equation} 
where the quantity $L$ is related to the cosmological constant. Thus the quantities $\mathcal{U}$, $\mathcal{V}$, and $\mathcal{W}$ for this metric are equal to \cite{Cotaescu:1998ay}, \cite{Cotaescu:2003be}
\begin{equation} \label{e: u v w de Sitter}
\mathcal{U} = 1, \quad \qquad \frac{\mathcal{V}}{y} = \frac{1}{L \sinh (\frac{y}{L})}, \quad \qquad  \mathcal{W} = \frac{1}{\cosh(\frac{y}{L})}.
\end{equation} 
Hence, taking into account the previous expressions, Eq.\ (\ref{e: Dirac matrix form}) simplifies to
\begin{equation} \label{e: Dirac matrix form Sitter}
\left( \begin{array}{cc} 
M \textrm{sech}(\tfrac{y}{L}) & \frac{{\rm d}}{{\rm d} y} +  \frac{\kappa}{L} \textrm{csch}(\tfrac{y}{L}) \\
\qquad & \qquad \\
- \frac{{\rm d}}{{\rm d} y} + \frac{\kappa}{L} \textrm{csch}(\tfrac{y}{L})  & - M \textrm{sech}(\tfrac{y}{L})
\end{array} \right)  \left( \begin{array}{c} f_{\omega, \kappa}^{(1)} (y)\\ \\ f_{\omega, \kappa}^{(2)}(y) \end{array} \right)
= \omega \left( \begin{array}{c} f_{\omega, \kappa}^{(1)}(y) \\ \\ f_{\omega, \kappa}^{(2)}(y) 
\end{array} \right).
\end{equation} 

Employing the coordinate $z=\tanh(y/L)$, related to the static radial coordinate $r$ by $r = z L$ \cite{Spradlin:2001pw}, we find that Eq.\ (\ref{e: Dirac matrix form Sitter}) transforms into
\begin{eqnarray} \label{e: radial z second Sitter}
\left[(1-z^2)^{\frac{1}{2}}\frac{{\rm d}}{{\rm d}z}  + \frac{\kappa}{z} \right] f_{\omega, \kappa}^{(2)}  &=&  \left[ \frac{ \tilde{\omega}}{(1-z^2)^{\frac{1}{2}}} - \tilde{M} \right]f_{\omega, \kappa}^{(1)} , \nonumber \\
\left[(1-z^2)^{\frac{1}{2}}\frac{{\rm d}}{{\rm d}z}  - \frac{\kappa}{z} \right] f_{\omega, \kappa}^{(1)}  &=&  - \left[ \frac{\tilde{\omega}}{(1-z^2)^{\frac{1}{2}}} + \tilde{M} \right]f_{\omega, \kappa}^{(2)} , 
\end{eqnarray}
where $\tilde{\omega} = \omega L $ and $\tilde{M} = M L$.

From Eqs.\ (\ref{e: radial z second Sitter}) and defining the new functions $R_1$ and $R_2$ by
\begin{equation}
R_1 = -i f_{\omega, \kappa}^{(1)} + f_{\omega, \kappa}^{(2)}, \qquad R_2 = -( i f_{\omega, \kappa}^{(1)} + f_{\omega, \kappa}^{(2)}),
\end{equation}  
we find that these functions satisfy
\begin{eqnarray} \label{e: radial z third Sitter}
\left[(1-z^2)^{\frac{1}{2}}\frac{{\rm d}}{{\rm d}z}  - \frac{ i\tilde{\omega}}{(1-z^2)^{\frac{1}{2}}}  \right] R_1  &=&  \left[ \frac{\kappa}{z}   - i \tilde{M} \right] R_2, \nonumber \\
\left[(1-z^2)^{\frac{1}{2}}\frac{{\rm d}}{{\rm d}z}  + \frac{ i  \tilde{\omega} } {(1-z^2)^{\frac{1}{2}}}  \right] R_2  &=&  \left[ \frac{\kappa}{z}  + i \tilde{M} \right] R_1 .
\end{eqnarray}
This system of differential equations has the same mathematical form already found in Refs.\ \cite{Lopez-Ortega:2006ig}, \cite{Lopez-Ortega:2004cq}, where we studied the massive Dirac field moving in $3D$ and $4D$ de Sitter spacetime.

Another method for simplifying the massless Dirac equation to a system of differential equations for the radial functions is given in Ref.\ \cite{Das:1996we} (see also \cite{Gibbons:1993hg}). The massive Dirac equation is not explicitly covered in Ref.\ \cite{Das:1996we}; therefore here we follow \cite{Cotaescu:2003be}. In Appendix \ref{Appendix 1} we show that if the background spacetime is the $D$-dimensional de Sitter, then the system of ordinary differential equations found by Das, \textit{et al.\ }in Ref.\ \cite{Das:1996we} can be transformed into Eqs.\ (\ref{e: radial z third Sitter}) with $\tilde{M} = 0$.

\section{Quasinormal frequencies of the Dirac field}
\label{Section 3}

In this section we find exact solutions to the system of differential equations (\ref{e: radial z third Sitter}) and using them we calculate the QN frequencies of the Dirac field propagating in $D$-dimensional de Sitter spacetime. We first find these QN frequencies in the massless case and then in the massive case. For the massless and massive Dirac equations we use different methods to make the simplification of the radial equations to hypergeometric type differential equations. The procedure for the massless Dirac field is simpler than that for the massive Dirac field, and we must use a more complex method to find the exact solutions in the last case.

\subsection{Massless Dirac field}
\label{Subsection 3.1}

Following a similar procedure to that used in Sect.\ 2 of Ref.\ \cite{Lopez-Ortega:2006ig} (see also \cite{Lopez-Ortega:2004cq}) we can find in a straightforward way the solutions to the system of differential equations (\ref{e: radial z third Sitter}) when $\tilde{M}$ is equal to zero. 

As in Ref.\ \cite{Lopez-Ortega:2004cq}, we make the ansatz for the radial functions $R_1$ and $R_2$ 
\begin{equation}
R_1 = v^{B_1} (1-v)^{C_1} \tilde{R}_1, \qquad R_2 = v^{B_2} (1-v)^{C_2}\tilde{R}_2,
\end{equation}
where
\begin{align}
& v=\frac{2z}{1+z},  \nonumber \\
& B_1 = B_2 = \pm \left( \frac{p}{2} + n\right), \nonumber \\
C_1 = & \left\{ \begin{array}{l} -\frac{i \tilde{\omega}}{2}  ,\\ \\ \frac{i \tilde{\omega}}{2} + \frac{1}{2} ,\end{array}\right. \qquad \quad
C_2 =  \left\{ \begin{array}{l} \frac{i \tilde{\omega}}{2}  ,\\ \\ \frac{1}{2} - \frac{i \tilde{\omega}}{2}  ,\end{array}\right.
\end{align} 
to find that the functions $\tilde{R}_1$ and $\tilde{R}_2$ must be solutions of the hypergeometric differential equations
\begin{equation} \label{e: hypergeometric differential}
v(1-v) \frac{{\rm d}^2 \tilde{R}_I}{{\rm d}v^2} + (c_I - (a_I +b_I + 1)v)\frac{{\rm d} \tilde{R}_I}{{\rm d}v} - a_I b_I \tilde{R}_I   = 0,
\end{equation} 
where $I = 1,2,$ and the quantities $a_I$, $b_I$, and $c_I$ are equal to
\begin{align} \label{e: a b c values Dirac}
&a_1 =  B_1 + C_1 + \frac{1}{2} - \frac{i \tilde{\omega}}{2}, \quad &a_2& = B_2 + C_2 + \frac{1}{2} + \frac{i \tilde{\omega}}{2}, \nonumber \\
&b_1 =  B_1 + C_1 + \frac{i \tilde{\omega}}{2}, \quad &b_2& = B_2 + C_2 - \frac{i \tilde{\omega}}{2}, \nonumber \\
&c_1 = 2 B_1 + 1, \quad &c_2 &= 2 B_2 + 1.
\end{align}

It is convenient to note that these solutions reduce to those already given in Refs.\ \cite{Lopez-Ortega:2006ig}, \cite{Lopez-Ortega:2004cq} for the massless Dirac field propagating in $3D$ and $4D$ de Sitter spacetime.

As noted in Refs.\ \cite{Lopez-Ortega:2006my}, \cite{Lopez-Ortega:2006ig}, \cite{Lopez-Ortega:2004cq} there are several possible values for the quantities $a_1$, $b_1$, and $c_1$ ($a_2$, $b_2$, and $c_2$) depending on the chosen values for $B_1$ and $C_1$ ($B_2$ and $C_2$). In the following we only study the case $B_1=\tfrac{p}{2} + n$ and $C_1=\tfrac{i \tilde{\omega}}{2} + \tfrac{1}{2}$ ($B_2=\tfrac{p}{2} + n$ and $C_2=\tfrac{i \tilde{\omega}}{2}$). In the other cases we expect to find identical physical results.

We first study the radial function $R_1$. As in Refs.\ \cite{Lopez-Ortega:2006my}, \cite{Lopez-Ortega:2006ig}, \cite{Lopez-Ortega:2004cq} we can show that the only solution of the differential equation (\ref{e: hypergeometric differential}), with $I=1$, that leads to a radial function regular at $r=0$ is the hypergeometric function
\begin{equation}
\tilde{R}_1 = {}_{2}F_{1}(a_1,b_1;c_1;v).
\end{equation}  
The other solution to the hypergeometric differential equation leads to a radial function divergent at $r=0$ (see Refs.\ \cite{Lopez-Ortega:2006my}, \cite{Lopez-Ortega:2006ig}, \cite{Lopez-Ortega:2004cq}). Thus
\begin{equation} \label{e: radial function regular}
R_1 = v^{n + p/2} (1-v)^{\tfrac{i \tilde{\omega}}{2} + \tfrac{1}{2}}\, {}_{2}F_{1}(a_1,b_1;c_1;v),
\end{equation} 
is regular at $r=0$.

As is well known, when $c_1-a_1-b_1$ is not an integer the hypergeometric function satisfies \cite{b:DE-books}  
\begin{align} \label{e: hypergeometric property z 1-z}
{}_2F_1(a_1,b_1;c_1;v) &= \frac{\Gamma(c_1) \Gamma(c_1-a_1-b_1)}{\Gamma(c_1-a_1) \Gamma(c_1 - b_1)} {}_2 F_1 (a_1,b_1;a_1+b_1+1-c_1;1-v) \nonumber \\
&+ \frac{\Gamma(c_1) \Gamma( a_1 + b_1 - c_1)}{\Gamma(a_1) \Gamma(b_1)} (1-v)^{c_1-a_1 -b_1}\nonumber \\ 
&\times {}_2F_1(c_1-a_1, c_1-b_1; c_1 + 1 -a_1-b_1; 1 -v),
\end{align} 
where $\Gamma(z)$ denotes the Gamma function. Exploiting this property of the hypergeometric function we can write the radial function (\ref{e: radial function regular}) as
\begin{align} \label{e: radial function expansion}
R_1 &= v^{n+p/2} \left\{ (1-v)^{\tfrac{i \tilde{\omega}}{2} + \tfrac{1}{4}} {}_2 F_1 (a_1,b_1;a_1+b_1+1-c_1;1-v)     \right. \nonumber \\ 
&\times \frac{\Gamma(c_1) \Gamma(c_1-a_1-b_1)}{\Gamma(c_1-a_1) \Gamma(c_1 - b_1)} +  \frac{\Gamma(c_1) \Gamma( a_1 + b_1 - c_1)}{\Gamma(a_1) \Gamma(b_1)} \nonumber \\ 
&\left. \times (1-v)^{- \tfrac{i \tilde{\omega}}{2} - \tfrac{1}{4}} {}_2F_1(c_1-a_1, c_1-b_1; c_1 + 1 -a_1-b_1; 1 -v) \right\}.
\end{align}

In the previous expression the first term in curly brackets represents an ingoing wave near the cosmological horizon, while the second term represents an outgoing wave. Hence, to get a purely outgoing wave near the cosmological horizon we must satisfy the condition 
\begin{equation} \label{e: conditions QNMs massless}
c_1 - a_1 = - n_2, \qquad\textrm{or}\qquad c_1-b_1=-n_2, \qquad n_2=0,1,2,\dots.
\end{equation} 
We cannot fulfill the first condition in Eqs.\ (\ref{e: conditions QNMs massless}), but from the second condition we obtain that the QN frequencies of the massless Dirac field moving in $D$-dimensional de Sitter background are equal to\footnote{As in Refs.\ \cite{Lopez-Ortega:2006my}, \cite{Lopez-Ortega:2006ig} we also call to the quantities $i \tilde{\omega}$ the de Sitter QN frequencies.}
\begin{equation} \label{e QN frequencies odd}
i \tilde{\omega} = n + \frac{p+1}{2} + n_2.
\end{equation} 

Now, we note that for these frequencies
\begin{equation}
c_1 - a_1 - b_1 = -\left(n+1+\frac{p}{2}+n_2 \right),
\end{equation} 
hence $c_1 - a_1 - b_1$ is not an integer for odd $p$, but for even $p$ this quantity is an integer. So for even $p$ the previous procedure does not produce the de Sitter QN frequencies of the massless Dirac field since in this case, if we impose the second condition of Eqs.\ (\ref{e: conditions QNMs massless}), then we contradict the assumption that the quantity $c_1-a_1-b_1$ is not an integer already used to expand the radial function in formula (\ref{e: radial function expansion})  \cite{Lopez-Ortega:2006my}, \cite{Lopez-Ortega:2006ig}. 

In Ref.\ \cite{Lopez-Ortega:2006ig} we noted that for a massless Dirac field moving in 3$D$ de Sitter spacetime the QNMs boundary conditions can be satisfied when $c_1-a_1-b_1$ is not an integer. Thus our result agrees with that of the previous reference.

For the massless Klein-Gordon field, the electromagnetic and gravitational perturbations we already showed that in even and odd dimensional spacetimes we cannot fulfill the boundary conditions for the de Sitter QNMs when the quantity $c_1 - a_1 - b_1$ is not an integer (see Sects.\ 2 and 3 of Ref.\ \cite{Lopez-Ortega:2006my} and Sect.\ 3.2 and Appendix B of Ref.\ \cite{Lopez-Ortega:2006ig}). For the massless Dirac field we cannot satisfy the boundary conditions when $c_1 - a_1 - b_1$ is not integer only in even dimensions. Therefore it shows a different behavior than the integer spin fields.
 
As in Refs.\ \cite{Lopez-Ortega:2006my}, \cite{Lopez-Ortega:2006ig}, for even dimensional de Sitter spacetimes we must analyze the case $c_1 - a_1 - b_1=-n_1$, $n_1=1,2,3,\dots$.\footnote{ For other integer values of $c_1-a_1-b_1$ we do not get de Sitter QN frequencies; see Refs.\ \cite{Lopez-Ortega:2006my}, \cite{Lopez-Ortega:2006ig} for more details.} For these values of $a_1$, $b_1$, and $c_1$ the hypergeometric function satisfies \cite{b:DE-books}
\begin{align} \label{e: hypergeometric property integer}
&{}_2F_1 (a_1,b_1;a_1+b_1-n_1;v) =   \frac{\Gamma(a_1+b_1-n_1) \Gamma(n_1)}{\Gamma(a_1)\Gamma(b_1)} (1-v)^{-n_1} \nonumber \\ 
&\times \sum_{s=0}^{n_1-1}\frac{(a_1-n_1)_s (b_1-n_1)_s}{s! (1-n_1)_s}(1-v)^s  - \frac{(-1)^{n_1} \Gamma(a_1+b_1-n_1)}{\Gamma(a_1-n_1)\Gamma(b_1-n_1)} \nonumber \\
&\times \sum_{s=0}^\infty \frac{(a_1)_s (b_1)_s}{s!(n_1+s)!}(1-v)^s [\textrm{ln}(1-v) -\psi(s+1) -\psi(s+n_1+1)  \nonumber \\
&  +\psi(a_1+s)+\psi(b_1+s)] ,
\end{align}  
where $\psi(z)={\rm d} \ln \Gamma(z)/{\rm d}z$, $(z)_0=1$, and $(z)_s=z(z+1)\cdots(z+s-1)$ for $s>1$. From the previous formula and the radial function (\ref{e: radial function regular}), to get a purely outgoing wave near the cosmological horizon we must impose the condition \cite{Lopez-Ortega:2006my}, \cite{Lopez-Ortega:2006ig} 
\begin{equation} \label{e: condition qnms even D}
a_1 - n_1 = -n_2, \qquad\textrm{or}\qquad b_1-n_1=-n_2.
\end{equation} 
We cannot fulfill the second condition in Eqs.\ (\ref{e: condition qnms even D}), but from the first condition we find that the QN frequencies of the massless Dirac field moving in $D$-dimensional de Sitter spacetime, for even $D$, are equal to
\begin{equation} \label{e QN frequencies even}
i \tilde{\omega} = n + \frac{p+1}{2} + n_2.
\end{equation} 

For the massless Dirac field propagating in $4D$ de Sitter spacetime in Sect.\ 3 of Ref.\ \cite{Lopez-Ortega:2006ig} we show that the QN frequencies are well defined. We also prove  that the QN frequencies can be calculated when the quantity $c_1-a_1-b_1$ is a negative integer. Hence our results agree with those for the massless Dirac field given in the previous reference.

Thus from formulas (\ref{e QN frequencies odd}) and (\ref{e QN frequencies even}) we can assert that for even and odd $D$ the de Sitter QN frequencies of the massless Dirac field are well defined. Analyzing the radial function $R_2$, in even and odd dimensions, we also find the de Sitter QN frequencies (\ref{e QN frequencies even}) (or (\ref{e QN frequencies odd})). It is convenient to notice that the quantities $i \tilde{\omega}$ of formula (\ref{e QN frequencies even}) are integers for odd $p$ and half-integers for even $p$, as the de Sitter QN frequencies of a massless Klein-Gordon coupled to curvature with coupling constant $\xi =(D-1)/(4D)$ (see expression (55) of \cite{Lopez-Ortega:2006my}). Nevertheless, if the mode number $n_2$ changes by the same quantity, then the increments in the imaginary part of the QN frequencies $\omega$ calculated here for the massless Dirac field and in Ref.\ \cite{Lopez-Ortega:2006my} for the massless scalar field are different.

Notice that the QN frequencies (\ref{e QN frequencies even}) reduce to those calculated in Sects.\ 2 and 3 of Ref.\ \cite{Lopez-Ortega:2006ig} for the massless Dirac field moving in 3$D$ and 4$D$ de Sitter spacetime (see also \cite{Du:2004jt}). We also point out that for two different odd dimensional de Sitter spacetimes the set of QN frequencies for the massless Dirac field are equal except for the first modes. This also happens for even dimensional de Sitter spacetimes.

\subsection{Massive Dirac field}

Exploiting a similar procedure to that given in Sect.\ 4 of Ref.\ \cite{Lopez-Ortega:2006ig} we can solve the system of differential equations (\ref{e: radial z third Sitter}) when $\tilde{M} \neq 0$. Thus in this section we find exact solutions to the equations of motion for a massive Dirac field propagating in $D$-dimensional de Sitter spacetime and we use them to calculate its QN frequencies.

Making the following ansatz for the radial functions $R_1$ and $R_2$ \cite{Lopez-Ortega:2006ig}
\begin{eqnarray} \label{e: ansatz R-1 R2 f-1 f-2}
R_1 = (1-z^2)^{-\frac{1}{4}} (1-z)^{\frac{1}{2}} (f_1 - f_2),  \nonumber \\
R_2 = (1-z^2)^{-\frac{1}{4}} (1+z)^{\frac{1}{2}} (f_1 + f_2),
\end{eqnarray} 
and taking $f_1$ and $f_2$ in the form
\begin{eqnarray} \label{csix:f1-f2-ansatz}
f_1 = x^{C_1} (1-x)^{B_1} \hat{f}_1, \qquad \qquad
f_2 = x^{C_2} (1-x)^{B_2} \hat{f}_2,
\end{eqnarray} 
where $x=z^2$ and 
\begin{align}
&\quad B_1 =  B_2 =\left\{ \begin{array}{l} \frac{i \tilde{\omega}}{2} + \frac{1}{4}, \\ \\ - \left( \frac{i \tilde{\omega}}{2} + \frac{1}{4} \right) ,\end{array}\right. \nonumber \\
C_1 = & \left\{ \begin{array}{l} \frac{\kappa}{2} ,\\  \\ \frac{1}{2}-\frac{\kappa}{2}, \end{array} \right. \qquad \qquad 
C_2 =  \left\{ \begin{array}{l} -\frac{ \kappa}{2}, \\  \\ \frac{1}{2}+\frac{\kappa}{2}, \end{array} \right. 
\end{align} 
we find that the functions $\hat{f}_1$ and $\hat{f}_2$ are solutions of the hypergeometric differential equations
\begin{equation} \label{e: hypergeometric differential x}
x(1-x) \frac{{\rm d}^2 \hat{f}_I}{{\rm d}x^2} + (c_I - (a_I +b_I + 1)x)\frac{{\rm d} \hat{f}_I}{{\rm d}x} - a_I b_I \hat{f}_I   = 0,
\end{equation} 
where the quantities $a_I$, $b_I$, and  $c_I$ are equal to
\begin{align} \label{e: a b c values Dirac massive}
&a_1 =  B_1 + C_1 + \frac{1}{4} + \frac{1}{2}\left(\frac{1}{2} + i \tilde{M}\right), &a_2& = B_2 + C_2 + \frac{1}{4} + \frac{1}{2}\left(\frac{1}{2} - i \tilde{M}\right), \nonumber \\
&b_1 =  B_1 + C_1 +  \frac{1}{4} - \frac{1}{2}\left(\frac{1}{2} + i \tilde{M}\right), 
&b_2& = B_2 + C_2 + \frac{1}{4} - \frac{1}{2}\left(\frac{1}{2} - i \tilde{M}\right), \nonumber \\
&c_1 = 2 C_1 + \frac{1}{2},  &c_2 &= 2 C_2 + \frac{1}{2}.
\end{align}

Depending on the chosen values for the quantities $B_I$ and $C_I$, we get several values for  $a_I$, $b_I$, and $c_I$ as already noted in Refs.\ \cite{Lopez-Ortega:2006my}, \cite{Lopez-Ortega:2006ig}. In the following we study the case  $B_1=\tfrac{i \tilde{\omega}}{2} + \tfrac{1}{4}$, $C_1= \tfrac{\kappa}{2}$, $B_2=\tfrac{i \tilde{\omega}}{2} + \tfrac{1}{4}  $, and $C_2= \tfrac{1}{2}+ \tfrac{\kappa}{2}$.

The solutions of Eqs.\ (\ref{e: hypergeometric differential x}) that lead to radial functions regular at $r=0$ are
\begin{equation} \label{e: f I}
\hat{f}_I = {}_{2}F_{1}(a_I,b_I;c_I;x).
\end{equation} 
The another linearly independent solution of Eq.\ (\ref{e: hypergeometric differential x}) leads to divergent radial functions at $r=0$. Furthermore, from Eqs.\ (\ref{e: radial z third Sitter}) we can show that if we choose $\hat{f}_1$ as in formula (\ref{e: f I}) then \cite{Lopez-Ortega:2006ig}
\begin{equation}
\hat{f}_2 = -\frac{b_1}{c_1} {}_{2}F_{1}(a_2,b_2;c_2;x),
\end{equation} 
and the radial function $R_1$ is equal to
\begin{equation} \label{e: radial massive Dirac}
R_1 = (1-z^2)^{B_1-\tfrac{1}{4}} (1-z)^{\tfrac{1}{2}}z^{2 C_1}\left[ {}_{2}F_{1}(a_1,b_1;c_1;z^2) +  \frac{b_1 z}{c_1} {}_{2}F_{1}(a_2,b_2;c_2;z^2)\right],
\end{equation} 
(a similar expression holds for the function $R_2$).

Using the property of the hypergeometric function given in formula (\ref{e: hypergeometric property z 1-z}) we can write the radial function $R_1$ (\ref{e: radial massive Dirac}) in a similar form to the complicated formula (92) of Ref.\ \cite{Lopez-Ortega:2006ig}. From this formula, to obtain a purely outgoing wave near the cosmological horizon we must satisfy the condition 
\begin{equation} \label{e: conditions frequency}
c_1-a_1+1=-n_2, \qquad\textrm{or}\qquad c_1-b_1=-n_2, \qquad n_2=0,1,2,\dots.
\end{equation} 
From the previous conditions and expressions (\ref{e: a b c values Dirac massive}), we find that the QN frequencies for the massive Dirac field moving in $D$-dimensional de Sitter spacetime are equal to
\begin{equation} \label{e: QN frequencies massive}
i \tilde{\omega} = n + \frac{p+3}{2} + 2 n_2 - i \tilde{M}, \qquad i \tilde{\omega} = n + \frac{p+1}{2} + 2 n_2 + i \tilde{M}.
\end{equation}  

Applying a similar method we can analyze the radial function $R_2$; in this case we also find the de Sitter QN frequencies (\ref{e: QN frequencies massive}). Taking the limit $\tilde{M}\to 0$ in expressions (\ref{e: QN frequencies massive}) we get the frequencies for the massless Dirac field given in formula (\ref{e QN frequencies even}). It is convenient to note that for the QN frequencies (\ref{e QN frequencies even}) and (\ref{e: QN frequencies massive}) the amplitude of the field diminishes with time. Also notice that the QN frequencies (\ref{e: QN frequencies massive}) have the same mathematical form than those for a coupled to curvature massive scalar field with coupling constant $\xi =(D-1)/(4D)$ which appear in formula (54) of Ref.\ \cite{Lopez-Ortega:2006my}.

In 4$D$ de Sitter spacetime the QN frequencies (\ref{e: QN frequencies massive}) take the form
\begin{equation} \label{e: QN 4D Sitter}ḉ
i \tilde{\omega} = n + \frac{5}{2} + 2 n_2 - i \tilde{M}, \qquad i \tilde{\omega} = n + \frac{3}{2} + 2 n_2 + i \tilde{M}.
\end{equation} 
The second set of quantities in formula (\ref{e: QN 4D Sitter}) is equal to the second set of QN frequencies given in expression (94) of Ref.\ \cite{Lopez-Ortega:2006ig}. To compare the other set of de Sitter QN frequencies, notice that in formula (94) of Ref.\ \cite{Lopez-Ortega:2006ig} we wrote the first set of the de Sitter QN frequencies as 
\begin{equation}
 i \tilde{\omega} = j + 2 n_1 - i  \tilde{M}, \qquad n_1=0,1,2,\dots,
\end{equation} 
($j$ is a half-integer, $j \geq \tfrac{1}{2}$), which is wrong, because in the previous reference we had imposed the condition (first condition in Eqs.\ (93) of Ref.\ \cite{Lopez-Ortega:2006ig})
\begin{equation}
 c_1-a_1=-n_2,
\end{equation} 
instead of the first condition in Eqs.\ (\ref{e: conditions frequency}). Hence the correct expression for these QN frequencies is
\begin{equation} \label{e: 4D corrected QN}
i \tilde{\omega} = j +2+ 2 n_1 - i  \tilde{M}.
\end{equation} 

The QN frequencies (\ref{e: 4D corrected QN}) coincide with those given in the first set of quantities in formulas (\ref{e: QN 4D Sitter}). Furthermore, in Ref.\ \cite{Lopez-Ortega:2006ig} we asserted that in 4$D$ de Sitter spacetime for each $j$, the frequencies $i \tilde{\omega} = j  - i  \tilde{M}$ are QN. From the corrected expression (\ref{e: 4D corrected QN}) for them, we infer that this affirmation is not true.

\section{Discussion}
\label{Section 4}

In this paper we found exact solutions to the radial equations of the Dirac field in $D$-dimensional de Sitter spacetime and using them we calculated its QN frequencies extending the results of Refs.\ \cite{Du:2004jt}, \cite{Lopez-Ortega:2006ig}. For the massless case we get well defined expressions for the QN frequencies in even and odd spacetime dimensions.

Moreover, our results for the de Sitter QN frequencies in the massless and massive cases reduce to those already calculated in Ref.\ \cite{Lopez-Ortega:2006my} for three and four dimensional spacetimes (see also Ref.\ \cite{Lopez-Ortega:2006ig}). Notice that for $D \geq 5$, the exact solutions to the Dirac equation and the values of the QN frequencies (\ref{e QN frequencies even}) and (\ref{e: QN frequencies massive}) have not been previously computed.

Using two different sets of solutions to the Dirac equation for the massless and massive cases,  we find the exact values of the de Sitter QN frequencies. For the massless Dirac field, we note that in odd dimensions the analysis is straightforward, while in even dimensional de Sitter spacetime, due to the non-standard arguments of the hypergeometric functions we must use more complex properties of these functions to calculate the de Sitter QNMs. Hence the massless Dirac field shows a different behavior than the integer spin fields already studied. 

Observe that taking the limit $M \to 0$ of the result obtained for the QN frequencies of the massive Dirac field we find the expression corresponding to the massless field. Also, notice that the result for the massless field was calculated employing a different set of exact solutions. 

From the results of Refs.\ \cite{Lopez-Ortega:2006my}, \cite{Lopez-Ortega:2006ig}, \cite{Lopez-Ortega:2004cq}, and this paper we can deduce that the massless Klein-Gordon field, the massless Dirac field, the electromagnetic (EM) field and the gravitational perturbations have well defined QN frequencies when they propagate in $D$-dimensional de Sitter spacetime. We also notice that for these massless fields the de Sitter QN frequencies are purely imaginary. For massive fields they can be complex. For other spacetimes in which the QN frequencies of some fields are purely imaginary see Refs.\ \cite{Lopez-Ortega:2006my}, \cite{Lopez-Ortega:2006ig}, \cite{Becar:2007hu}.

Summarizing the findings of Refs.\ \cite{Lopez-Ortega:2006my}, \cite{Lopez-Ortega:2006ig}, \cite{Lopez-Ortega:2004cq} and the previous sections we present the QN frequencies $\omega$ for the $D$-dimensional de Sitter spacetime in Table 1. In this table $n$ is an integer, $n_2=0,1,2,\dots$, $\tilde{M}=M L$, $M$ being the mass of the field, $q=0,1$, and $2$ for tensor, vector, and scalar gravitational perturbations, respectively.

\begin{table}[t]
\caption{de Sitter QN frequencies }
\centering
\begin{tabular}{ll}
\hline\noalign{\smallskip}
 Field & QN frequency $\omega$ \\[3pt]
\hline\noalign{\smallskip}
Massive scalar field  & $-i\left( n+2n_2+\tfrac{p+1}{2}+[(\tfrac{p+1}{2})^2-\tilde{M}^2 ]^{1/2}\right)/L$ \\
($n\geq 0$) &  $-i\left( n+2n_2+\tfrac{p+1}{2}-[(\tfrac{p+1}{2})^2-\tilde{M}^2 ]^{1/2}\right)/L$ \\ \noalign{\smallskip}\hline \noalign{\smallskip}
Massive Dirac field & $-i (n+\tfrac{p+3}{2}+2n_2-i\tilde{M})/L$  \\
($n\geq 0$)  & $-i (n+\tfrac{p+1}{2}+2n_2+i\tilde{M})/L$ \\ \noalign{\smallskip} \hline \noalign{\smallskip}
EM field modes \textbf{I} & $-i(n+p-1+2n_2)/L$ \\
($n\geq 1$) & $-i(n+2+2n_2)/L$  \\ \noalign{\smallskip} \hline \noalign{\smallskip}
EM field modes \textbf{II} & $-i(n+p+2n_2)/L$ \\
($n\geq 1$) & $-i(n+1+2n_2)/L$ \\ \noalign{\smallskip} \hline \noalign{\smallskip}
Gravitational perturbations & $-i(n+p+1-q+2n_2)/L$ \\
($n\geq 2$) & $-i(n+q+2n_2)/L$ \\ \noalign{\smallskip} \hline \noalign{\smallskip}
\end{tabular}
\end{table}

To prove or disprove our results it would be interesting to obtain numerical values for the de Sitter QN frequencies. To our knowledge, for the Rarita-Schwinger and Proca fields propagating in $D$-dimensional de Sitter background, a similar analysis to that of Refs.\ \cite{Lopez-Ortega:2006my}, \cite{Lopez-Ortega:2006ig}, \cite{Lopez-Ortega:2004cq}, and the present paper is missing.

Is there any relation between the de Sitter QNMs and the thermodynamics properties of the cosmological horizon? We believe that this fact should be investigated additionally. For the case of a static black hole, see Refs.\ \cite{Hod:1998vk} for more details.


\section{Acknowledgements}

I thank Dr.\ M.\ A.\ P\'erez Ang\'on for his interest in this paper and also for proofreading the manuscript. This work was supported by CONACyT and SNI (M\'exico).


\begin{appendix}

\section{Das-Gibbons-Mathur equations}
\label{Appendix 1}

The metric of a $D$-dimensional spherically symmetric spacetime in isotropic coordinates takes the form\footnote{For consistency with Ref.\ \cite{Das:1996we} in this Appendix we use the signature $(-++\dots+)$ for the spacetime metric.}
\begin{equation} \label{e:metric-general}
{\rm d}s^2= - f(\bar{r})\, {\rm d}t^2 + g(\bar{r})\,[{\rm d} \bar{r}^2 +\bar{r}^2 {\rm d} \Omega^2_p],
\end{equation} 
where the variable $\bar{r}$ is called isotropic radial coordinate and $D=p+2$ as in Sects.\ \ref{Section 2} and \ref{Section 3}.

For a $D$-dimensional spherically symmetric background whose metric is written using the isotropic coordinates of formula (\ref{e:metric-general}) in Ref.\ \cite{Das:1996we} Das, \textit{et al.\ }showed that the equations of motion for a massless Dirac field reduce to the system of ordinary  differential equations (Eqs.\ (24) of Ref.\ \cite{Das:1996we})
\begin{align} \label{e: Das radial equations}
h(\bar{r}) \left[ \frac{{\rm d}}{{\rm d} \bar{r}} + \frac{p + n}{\bar{r}} \right] G_n^1 & = i \omega F_n^1 ,\nonumber\\
h(\bar{r}) \left[ \frac{{\rm d}}{{\rm d} \bar{r}} - \frac{n}{\bar{r}} \right]F_n^1 & = i \omega G_n^1,
\end{align}
where $h = \sqrt{f/g}$, $F_n^1$, and $G_n^1$ are functions of $\bar{r}$ and $n=0,1,2,\dots$. For more details see Ref.\ \cite{Das:1996we}.

Using the relation
\begin{equation}
 \bar{r} = \frac{L + \sqrt{L^2 -r^2}}{r},
\end{equation} 
that exists between isotropic and static coordinates for the de Sitter spacetime, we find that the system of ordinary differential equations (\ref{e: Das radial equations}) becomes
\begin{eqnarray} \label{e: radial z simplified}
\left[(1-z^2)^{\frac{1}{2}}\frac{{\rm d}}{{\rm d}z}  - \frac{p + n}{z} \right] G_n^1 = - \frac{i \tilde{\omega}}{(1-z^2)^{\frac{1}{2}}} F_n^1 , \nonumber \\
\left[(1-z^2)^{\frac{1}{2}}\frac{{\rm d}}{{\rm d}z}  + \frac{n}{z} \right] F_n^1 = - \frac{i \tilde{\omega}}{(1-z^2)^{\frac{1}{2}}} G_n^1 ,
\end{eqnarray} 
where $r = zL $ and $\tilde{\omega} = \omega L $ as in Sect.\ \ref{Section 2}.

Replacing in Eqs.\ (\ref{e: radial z simplified}) $G_n^1 \to - G_n$, $F_n^1 \to F_n$, and taking 
\begin{equation}
 G_n = H \tilde{G}_n, \qquad \qquad F_n = H \tilde{F}_n,
\end{equation} 
where 
\begin{equation}
 H = \left( \frac{1 +  \sqrt{1-z^2}}{z}\right)^{- p/2},
\end{equation} 
we get that the functions $\tilde{G}_n$ and $\tilde{F}_n$ satisfy
\begin{eqnarray} \label{e: radial z second}
\left((1-z^2)^{\frac{1}{2}}\frac{{\rm d}}{{\rm d}z}  - \left( n + \frac{p}{2}\right) \frac{1}{z} \right) \tilde{G}_n =  \frac{i \tilde{\omega}}{(1-z^2)^{\frac{1}{2}}} \tilde{F}_n , \nonumber \\
\left((1-z^2)^{\frac{1}{2}}\frac{{\rm d}}{{\rm d}z}  + \left( n + \frac{p}{2}\right) \frac{1}{z} \right) \tilde{F}_n =  \frac{i \tilde{\omega}}{(1-z^2)^{\frac{1}{2}}} \tilde{G}_n . 
\end{eqnarray}

Defining the new radial functions
\begin{equation}
 R_1 = \tilde{G}_n + \tilde{F}_n, \qquad \qquad R_2 = \tilde{G}_n - \tilde{F}_n,
\end{equation} 
we obtain that the system of differential equations (\ref{e: radial z second}) becomes
\begin{eqnarray} \label{e: radial z third}
\left((1-z^2)^{\frac{1}{2}}\frac{{\rm d}}{{\rm d}z}  - \frac{i \tilde{\omega}}{(1-z^2)^{\frac{1}{2}}} \right) R_1 = \left( n + \frac{p}{2}\right) \frac{1}{z} R_2, \nonumber \\
\left((1-z^2)^{\frac{1}{2}}\frac{{\rm d}}{{\rm d}z}  + \frac{i \tilde{\omega}}{(1-z^2)^{\frac{1}{2}}} \right) R_2 = \left( n + \frac{p}{2}\right) \frac{1}{z} R_1. 
\end{eqnarray}
Thus in $D$-dimensional de Sitter background, the system of differential equations (\ref{e: Das radial equations}) for the massless Dirac field simplify to Eqs.\ (\ref{e: radial z third Sitter}) with $\tilde{M} = 0$ and $\kappa = \frac{p}{2} + n$. We already solved this system of differential equations in Sect.\ \ref{Section 3}.

\section{Effective potentials}
\label{Appendix 2}

For several spacetimes in which exact solutions to the equations of motion for the classical fields can be found, in Appendix C of Ref.\ \cite{Lopez-Ortega:2006my} we noted that the effective potentials of the Schr\"odinger type equations are of the P\"oschl-Teller form\footnote{Here the variable $x$ is different from that of Sect.\ \ref{Section 3}.} 
\begin{equation} \label{e: Teller potential}
 V(x) = \frac{{\mathcal A}}{\sinh^2(x)} + \frac{{\mathcal B}}{\cosh^2(x)}.
\end{equation} 
We presented the values of ${\mathcal A}$ and ${\mathcal B}$ for these spacetimes in Table 1 of Ref.\ \cite{Lopez-Ortega:2006my}.

Applying the method described in Ref.\ \cite{Chandrasekhar book}, we can prove that Eqs.\ (\ref{e: radial z third Sitter}) for the massive Dirac field in de Sitter spacetime reduce to a pair of Schr\"odinger type equations with effective potentials
\begin{equation} \label{e: effective potentials}
V_{\pm} = W^2 \pm \frac{{\rm d} W}{{\rm d} \hat{r}_*},
\end{equation} 
where
\begin{align}
W&=\frac{ (1-z^2)^{\tfrac{1}{2}} [\tilde{\kappa}^2 + (Mz)^2]^{\tfrac{3}{2}} }{ z [\tilde{\kappa}^2 + (Mz)^2] + \frac{\tilde{\kappa} M z (1-z^2)}{2 \omega L}}, \nonumber \\
\hat{r}_*&=r_* + \frac{1}{2\omega} \arctan\left(\frac{M z }{\tilde{\kappa}} \right), \nonumber \\
\tilde{\kappa} &= \frac{\kappa}{L},
\end{align} 
$z$ is defined in Sect.\ \ref{Section 2}, $r_*$ is the usual tortoise coordinate for the $D$-di\-men\-sion\-al de Sitter spacetime and the quantity $W$ is known as a superpotential for the potentials $V_{\pm}$ \cite{Dutt:1987va}.

For the massless field, we can write the effective potentials (\ref{e: effective potentials}) as
\begin{equation} \label{e: Dirac potentials}
V_{\pm} = \frac{\tilde{\kappa}^2}{\sinh^2(\tfrac{r_*}{L})} \mp \frac{\tilde{\kappa} \cosh(\tfrac{r_*}{L})}{L \sinh^2(\tfrac{r_*}{L})}.
\end{equation} 
Therefore for a massless Dirac field moving in $D$-di\-men\-sion\-al de Sitter spacetime, in these variables the effective potentials (\ref{e: Dirac potentials}) are not of P\"oschl-Teller type (\ref{e: Teller potential}). Nevertheless, we note that the potentials (\ref{e: Dirac potentials}) are of Rosen-Morse type (see Table I in Ref.\ \cite{Dutt:1987va}). This fact has not been previously observed.\footnote{For the massless Dirac field propagating in the BTZ black hole we can simplify the equations of motion to Schr\"odinger type equations with effective potentials equal to (see expression (74) in \cite{Lopez-Ortega:2006my} and Ref.\ \cite{Cardoso:2001hn})
\begin{displaymath}
V_{\pm} = \frac{l^2/L^2}{\cosh^2\left(\frac{M^{1/2}}{L}r_*\right)} \pm \frac{\frac{l M^{1/2}}{L^2} \sinh \left(\frac{M^{1/2}}{L}r_*\right) }{\cosh^2\left(\frac{M^{1/2}}{L}r_*\right)},
\end{displaymath} 
which are of Morse type \cite{Dutt:1987va}. Furthermore, observe that the P\"oschl-Teller, Rosen-Morse, and Morse  potentials are all shape invariant \cite{Dutt:1987va}.}

Furthermore, the quantity 
\begin{equation}
W = \tilde{\kappa} \frac{(1-z^2)^{1/2}}{z} = \tilde{\kappa}\,\, \textrm{csch} \left(\frac{r_*}{L}\right),
\end{equation} 
is a superpotential for the potentials (\ref{e: Dirac potentials}). Thus we can expect that the potentials (\ref{e: Dirac potentials}) have the same spectrum \cite{Dutt:1987va}, as we already noted.

In Ref.\ \cite{Lopez-Ortega:2006my} we showed that the modes {\bf I} and {\bf II} of the electromagnetic field have a different spectrum of the de Sitter QN frequencies (for fixed angular momentum number $l$, the QN frequencies have different parity in odd dimensions, while in even dimensions a finite number of QN frequencies, which depend on the spacetime dimension, are different).

For the electromagnetic field, the effective potentials of the modes $\textbf{I}$ and $\textbf{II}$ in $D$-dimensional de Sitter spacetime are equal to (Table 1 of Ref.\ \cite{Lopez-Ortega:2006my})
\begin{eqnarray} \label{e: potentials em}
V_{\textbf I} = \frac{l(l+D-3) +\tfrac{(D-2)(D-4)}{4}}{\sinh^2(x)} - \frac{\frac{(D-4)(D-6)}{4}}{\cosh^2(x)}, \\
V_\textbf{II} = \frac{l(l+D-3) +\tfrac{(D-2)(D-4)}{4}}{\sinh^2(x)} - \frac{\frac{(D-2)(D-4)}{4}}{\cosh^2(x)},
\end{eqnarray} 
with $x=r_*/L$. Hence we can write the Schr\"odinger type equation in the form \cite{Lopez-Ortega:2006my}
\begin{equation}
\left( \frac{{\rm d}^2}{{\rm d}x^2} + (\tilde{\omega}^2 + \Omega) - \tilde{V}_\textbf{I,II} \right)\Phi_\textbf{I,II} = 0,
\end{equation} 
where 
\begin{equation}
\tilde{V}_\textbf{I,II} = V_\textbf{I,II} + \Omega,
\end{equation} 
and
\begin{equation}
\Omega = (D-4+l)^2.
\end{equation} 

Furthermore, we find that $\tilde{V}_\textbf{II}$ can be written in the form (see formula (\ref{e: effective potentials}))
\begin{equation} \label{e: supersymetric relation}
\tilde{V}_\textbf{II} = - \frac{{\rm d}W_\textbf{II}}{{\rm d}x} + W_\textbf{II}^2,
\end{equation} 
where
\begin{equation}
W_\textbf{II} = \left( \frac{D}{2} -2 \right) \tanh(x) + \left( \frac{D}{2} + l -2 \right) \coth(x).
\end{equation} 
Therefore the SUSY related potential to $\tilde{V}_\textbf{II}$ is \cite{Dutt:1987va} 
\begin{align} \label{e: SUSY partner II}
\tilde{V}^{S}_\textbf{II} &= \frac{{\rm d} W_\textbf{II}}{{\rm d}x} + W_\textbf{II}^2 \nonumber \\
& = \frac{(l-1)(l+D-4) +\tfrac{(D-2)(D-4)}{4}}{\sinh^2(x)} - \frac{\frac{(D-4)(D-6)}{4}}{\cosh^2(x)} + (D+l-4)^2,
\end{align} 
which is not equal to $\tilde{V}_{\textbf I}$.

Thus the potentials $\tilde{V}_\textbf{I}$ and $\tilde{V}_\textbf{II}$ are not supersymmetric partner potentials. An interesting question is whether the modes \textbf{I} and \textbf{II} have a different spectrum of the de Sitter QN frequencies because their effective potentials are not supersymmetric partners.

\section{de Sitter QNMs, an alternative method of computation}
\label{Appendix 3}

In Refs.\ \cite{Lopez-Ortega:2006my}, \cite{Lopez-Ortega:2006ig} we calculate the de Sitter QN frequencies of several massless fields with integer spin. Here we present a slightly different method of computation for these frequencies. We shall study in detail the case of the vector gravitational perturbations propagating in $D$-dimensional de Sitter background and at the end of this section we comment on the results obtained for the other massless fields.

In de Sitter spacetime the radial function $R(z)$ for the vector type gravitational perturbations satisfies the differential equation \cite{Natario:2004jd}, \cite{Lopez-Ortega:2006my}
\begin{equation} \label{e: radial vector type}
(1-z^2) \frac{{\rm d}^2 R(z) }{{\rm d}z^2} -2z\frac{{\rm d} R(z)}{{\rm d}z} + \left( \frac{\tilde{\omega}^2}{1-z^2} - \frac{B(B+1)}{z^2} + A \right) R(z) = 0, 
\end{equation} 
where $A=p(p-2)/4$, $B=(2n+p-2)/2$, $n$ an integer $\geq 2$, and $z$ was already defined.

Following Leaver \cite{Leaver:1985ax} we take $R(z)$ in the form
\begin{equation} \label{e: ansatz radial vector}
R(z)=z^{B+1}(1-z^2)^{-i \tilde{\omega}/2} R_1(z),
\end{equation} 
that fulfills the boundary conditions for the de Sitter QNMs if the function $R_1(z)$ satisfies some conditions (see below), since $z^{B+1}$ is a regular function at $z=0$ and $(1-z^2)^{-i \tilde{\omega} /2 }$ represents an outgoing wave near the cosmological horizon. Substituting expression (\ref{e: ansatz radial vector}) into Eq.\ (\ref{e: radial vector type}) we find that the function $R_1(z)$ must be a solution to
\begin{align}
&\frac{{\rm d}^2 R_1(z) }{{\rm d}z^2} + \left( \frac{1-i \tilde{\omega}}{z+1} + \frac{2B+2}{z} + \frac{1-i \tilde{\omega}}{z-1} \right) \frac{{\rm d} R_1(z)}{{\rm d}z} +\left( \frac{3 i \tilde{\omega} + 2 i \tilde{\omega} B + A + \tilde{\omega}^2 }{2(z+1)} \right. \nonumber \\
& \left. - \frac{2 + 3 B + B^2}{2(z+1)} + \frac{ 3 B + B^2 -3 i \tilde{\omega} - 2 i \tilde{\omega} B - A - \tilde{\omega}^2 + 2 }{2(z-1)} \right) R_1(z) = 0.
\end{align} 

Making the change of variable $x=1-z^2$ we get that the function $R_1(x)$ satisfies the ordinary differential equation
\begin{align}
x(1-x)\frac{{\rm d}^2 R_1(x) }{{\rm d}x^2} + \left( 1-i \tilde{\omega} + \left( i \tilde{\omega}  - n -\frac{p}{2} - \frac{3}{2} \right)x \right)\frac{{\rm d} R_1(x)}{{\rm d}x} \nonumber \\
+\left( \frac{i \tilde{\omega} ( 1 + 2 n + p) + \tilde{\omega}^2 - (n + p)(n+1) }{4} \right) R_1(x) = 0,
\end{align} 
whose solution is \cite{b:DE-books} 
\begin{align} \label{e: solution radial x variable}
R_1(x) =&  C_1 \, {}_{2}F_{1}(\frac{n-i \tilde{\omega} + p}{2},\frac{n+1-i \tilde{\omega}}{2}; 1-i\tilde{\omega} ; x) \nonumber \\
& + C_2 \, x^{i \tilde{\omega}} {}_{2}F_{1}(\frac{n+ i \tilde{\omega} + p}{2},\frac{n+1+i \tilde{\omega}}{2}; 1 + i\tilde{\omega} ; x),
\end{align}
where $C_1$ and $C_2$ are constants.

We note that the second term in formula (\ref{e: solution radial x variable}) produces an ingoing wave near the cosmological horizon, therefore we take $C_2 = 0$. Thus to satisfy the boundary conditions $R_1(x)$ must be equal to
\begin{equation}
R_1(x) = C_1 \, {}_{2}F_{1}(\frac{n-i \tilde{\omega} + p}{2},\frac{n+1-i \tilde{\omega}}{2}; 1-i\tilde{\omega} ; x) . 
\end{equation}  
Also, notice that the parameters of the previous hypergeometric function satisfy
\begin{equation}
c-a-b= \frac{1}{2} - \frac{p}{2} -n.
\end{equation} 
Consequently the function $R_1(x)$ is divergent at $x=1$ ($z=0$) and this divergence leads to a radial function divergent at the origin, except when $R_1(x)$ is a polynomial \cite{b:DE-books}. Therefore in order that the radial function $R(z)$ be regular at $z=0$, we must satisfy the condition \cite{b:DE-books}
\begin{equation}
\frac{1 - i \tilde{\omega} + n }{2} = -n_2, \qquad \textrm{or} \qquad \frac{n - i \tilde{\omega} + p }{2} = -n_2, \qquad \qquad n_2=0,1,2,\dots
\end{equation} 

Hence the de Sitter QN frequencies of the vector type gravitational perturbations take the form
\begin{equation}
i \tilde{\omega} = n + 1 +2n_2, \qquad \qquad i \tilde{\omega} = n + p +2n_2,
\end{equation} 
which are equal to those given in expression (37) of Ref.\ \cite{Lopez-Ortega:2006my}. Furthermore using the previous method we compute the de Sitter QN frequencies for the tensor and scalar gravitational perturbations, the electromagnetic modes {\bf I} and {\bf II}, and the massless Klein-Gordon field. The results that we get are the de Sitter QN frequencies already calculated in Ref.\ \cite{Lopez-Ortega:2006my}. For the massive Klein-Gordon field we also found the QN frequencies computed in the previous reference.

\end{appendix}

\end{document}